\documentstyle[aps,manuscript,epsfig]{revtex}

\def\D{\displaystyle}
\begin{document}
\title{Electromagnetic wave scattering from conducting self-affine surfaces :
       An analytic and numerical study}
\author{Ingve Simonsen$^{*\dag}$, Damien Vandembroucq$^{*}$
 and St\'ephane Roux$^{*}$}
\address{
$^*$Laboratoire CNRS/Saint-Gobain ``Surface du Verre et Interfaces'', 
    93303 Aubervilliers Cedex, France\\
$^\dag$Department of Physics, 
       The Norwegian University of Science and Technology,
       N-7491 Trondheim, Norway}
\date{May 15, 2000}
\maketitle

\begin{abstract}
We derive an analytical expression for the scattering of a scalar wave
from a perfectly conducting self-affine one dimensional surface in the
framework of the Kirchhoff approximation.  We show that most of the
results can be recovered {\it via} a scaling analysis.  We identify
the typical slope taken over one wavelength as the relevant parameter
controlling the scattering process. We compare our predictions with
direct numerical simulations performed on surfaces of varying
roughness parameters and confirm the broad range of applicability of
our description up to very large roughness.  Finally we check that a
non zero electrical resistivity provided small does not invalidate our
results.
\end{abstract}

\section{Introduction}

Although studied for more than fifty years\cite{Rayleigh} wave
scattering from rough surfaces remains a very active field. This
constant interest comes obviously from the broad variety of its
applications domains which include remote sensing, radar technology,
long range radio-astronomy, surface physics, {\it etc.}, but from the
fundamental point of view, the subject has also shown a great vitality
in recent years. One may particularly cite the backscattering
phenomena originating either from direct multiple scattering
\cite{backscattering1,backscattering2} or mediated by surface plasmon
polaritons \cite{SPP1,SPP2,SPP3,SPP4,SPP5}. Remaining in the context
of single scattering a large amount of works have also been devoted to
the development of reliable analytical approximations
\cite{MFT,Thorsos,Ogilvy,Book:Beckmann}.  In all cases, the efficency
of any analyitical approximation relies on a proper description of the
surface roughness. In most models the height statistics are assumed to
be Gaussian correlated.  In this paper we address the question of wave
scattering from rough self-affine metallic surfaces. Since the
publication of the book by B.\ B.\ Mandelbrot about {\it the fractal
  geometry of nature}\cite{Mandelbrot}, scale invariance has become a
classical tool in the description of physical objects. In the more
restricted context of rough surfaces, scale invariance takes the form
of self-affinity.  Classical examples of rough surfaces obeying this
type of symmetry are surfaces obtained by fracture\cite{EB97} or
deposition\cite{Meakin}. More recently it was shown that cold rolled
aluminum surfaces\cite{laminage} could also be successfully described
by this formalism. When dealing with wave scattering from rough
surfaces, this scale invariance has one major consequence of interest,
it is responsible for long range correlations. After early works by
Berry\cite{Berry}, lots of works have been performed to study the
effects of fractal surfaces on wave scattering. Most of these works
were numerical (see for example
Refs.~\cite{Jaggard,Shepard,MacSharry,Lin,Chen,Sheppard,Sanchez,Sanchez2,Zhao})
and very few analytical or experimental results have been published.
Notable exceptions are due to Jakeman and his
collaborators\cite{Jakeman86,Jakeman88} who worked on 
diffraction through self-affine phase screens in the eighties and more
recent works applied to the characterization of growth
surfaces\cite{YangPRB93,ZhaoPRB97,ZhaoJAP98}. We recently gave a
complete analytical solution to the problem of wave scattering from a
perfectly conducting self-affine surface\cite{SRVPRE} in the Kirchhoff
approximation. In the following we present a complete derivation of
this expression and we deduce from it analytical expressions for the
width of the specular peak and the diffuse tail. These results are
compared to direct numerical simulations. We show evidence that the
crucial quantitative parameter is the slope of the surface taken over
one wavelength.


\section{The Scattering System}

\begin{figure}[t!]
     \begin{center}
       \epsfig{file=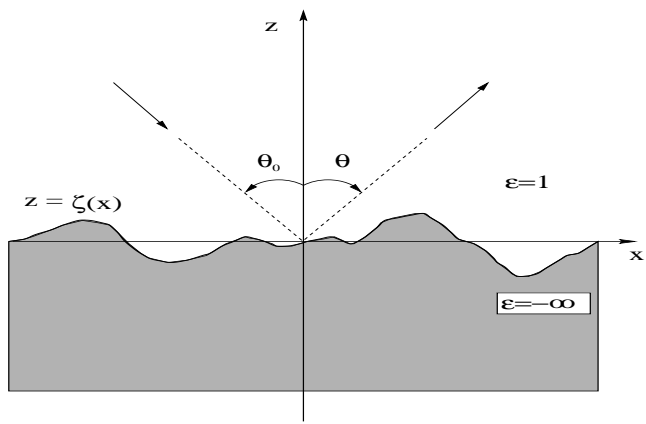,width=8.5cm,height=5.5cm}
     \end{center}
     \caption{The scattering geometry considered in this paper.}
     \label{Fig:Geometry}
 \end{figure}

The scattering system considered in the present work is depicted in
Fig.~\ref{Fig:Geometry}. It consists of vacuum in the region
$z>\zeta(x)$ and a perfect conductor in the region $z<\zeta(x)$.  The
incident plane is assumed to be the $xz$-plane. This system is
illuminated from the vacuum side by an $s$-polarized plane wave of
frequency $\omega=2\pi/\lambda$.  The angles of incidence and
scattering respectively are denoted by $\theta_0$ and $\theta$, and
they are defined positive according to the convention indicated in
Fig.~\ref{Fig:Geometry}.

In this paper we will be concerned with $1+1$-dimensional self-affine
surfaces $z=\zeta(x)$.  A surface is said to be self-affine between
the scales $\xi_{-}$ and $\xi_{+}$, if it remains (either exactly or
statistically) invariant in this region under transformations of the
form:
\begin{mathletters}
  \label{scaling-rel}
  \begin{eqnarray}
     \Delta x         &\to&   \mu \Delta x, \\
     \Delta \zeta     &\to&   \mu^{H} \Delta \zeta,
   \end{eqnarray}
\end{mathletters}

for all positive real numbers $\mu$. Here $H$ is the roughness
exponent, also known as the Hurst exponent, and it characterizes this
invariance.  This exponent is usually found in the range from zero to
one.  A statistical translation of the previous statement is that the
probability $p(\Delta \zeta;\Delta x)$ of having a height difference
in the range $[\Delta \zeta,\Delta \zeta+d\Delta \zeta]$ over the (lateral) distance $\Delta x$
is such that:
\begin{eqnarray}
p\left(\Delta \zeta ; \Delta x \right) d\Delta \zeta
     = p\left(\mu^{H} \Delta \zeta; \mu \Delta x \right) d\mu^{H} \Delta \zeta. 
\end{eqnarray}
Simple algebra based on the scaling relation~(\ref{scaling-rel}) gives
that the standard deviation of the height differences $\zeta(x+\Delta
x)-\zeta(x)$ measured over a window of size $\Delta x$ can be written
as
\begin{mathletters}
  \label{Eq:sigma}
  \begin{eqnarray}
    \label{Eq:sigma-a}
    \sigma(\Delta x) &=& \ell^{1-H} \Delta x^H,
  \end{eqnarray}
  and the (mean) slope of the surface as
   \begin{eqnarray}
    s(\Delta x) &=& \left(\frac{\ell}{\Delta x} \right)^{1-H}.
  \end{eqnarray}
\end{mathletters}
In these equations $\ell$ denotes a length scale known as the
topothesy. It is defined as $\sigma(\ell)=\ell$ (or $s(\ell)=1$). 

Alternatively, Eq.~(\ref{Eq:sigma-a}) can be written in the form
\begin{eqnarray}
   \sigma \left( \Delta x \right) = \sigma(\lambda)\left( \frac{\Delta x}{\lambda} \right)^{H}
= \lambda s(\lambda)\left( \frac{\Delta x}{\lambda} \right)^{H}
\;,
\end{eqnarray}
where we use the wavelength $\lambda$ of the scattering problem as a
normalization length. Here $\sigma(\lambda)$ and $s(\lambda)$ are
respectively the typical height difference and slope over one
wavelength as defined by Eqs.~(\ref{Eq:sigma}). Note that we could
have used any length scale for the normalization, like for instance
the topothesy. However, the choice made here was dictated by the
physical problem studied.  Using similar scaling arguments one can
show that the power density function of the height profile ${\cal
  P}(k)$ depends on the wave number $k$ as a power law:
\begin{equation}
\label{SA-power-spec}
{\cal P}(k)= \left| 
            \int_{-\infty}^{\infty} \zeta(x) \exp (ikx) dx 
       \right|^2 \propto k^{-1-2H}.
\end{equation}

In the case of a Gaussian height distribution, the probability
  $p(\Delta \zeta;\Delta x)$ reads:
\begin{eqnarray}
\label{dist}
p(\Delta \zeta;\Delta x)= \frac{\lambda^{H-1}}{\sqrt{2\pi} s(\lambda) \Delta x^{H}}
\exp \left[ -\frac{1}{2} \left( 
\frac{\lambda^{H-1} \Delta \zeta} { s(\lambda) \Delta x^{H}}
\right)^2 \right].
\end{eqnarray}
The self-affine profile is thus fully characterized by the roughness
exponent $H$, the slope $s(\lambda)$ (which is nothing but an
amplitude parameter) and the bounds of the self-affine regime
$\xi_{-}$ and $\xi_{+}$.

Numerous methods have been developed to estimate these parameters (see
for example Ref. \cite{Schmittbuhl}), most of them use the expected
power law variation of a roughness estimator computed over spatial
ranges of varying size. This roughness estimator can be a height
standard deviation, the difference between the maximum and the minimum
height, {\it etc.} It is also classical to use directly the power density
function of the profile. More recently the wavelet analysis has beeen
shown to offer a very efficient method to compute the roughness
exponent of self-affine surfaces~\cite{ingvewlt}.

\section{Scattering Theory}

In the following we consider the scattering of $s$-polarized
electromagnetic waves from a one-dimensional, random, Gaussian
self-affine surface $z=\zeta(x)$.  It will be assumed that the lower
limit of the self-affine regime $\xi_-$ is smaller than the
wavelength, $\lambda$, of the incident wave.  
For the present scattering system, where the roughness is
one-dimensional, the complexity of the problem is reduced
significantly. The reason being that there is no depolarization and
therefore the original three-dimensional vector scattering problem
reduces to a two-dimensional scalar problem for the single
non-vanishing 2nd component for the electric field,
$\Phi(x,z|\omega)=E_y(x,z|\omega)$, which should satisfy the (scalar)
Helmholtz equation
\begin{eqnarray}
    \label{Wave-eq}
    \left(\partial^2_x + \partial^2_z + \frac{\omega^2}{c^2}
    \right)\Phi(x,z|\omega) &=& 0,
\end{eqnarray}
with vanishing boundary condition on the randomly rough surface
$z=\zeta(x)$, and outgoing wave condition at infinity.  In the far
field region, above the surface, the field can be represented as the
sum of an incident wave and scattered waves:
\begin{eqnarray}
    \label{asymp-form}
    \Phi(x,z|\omega) &=& \Phi_0(x,z|\omega)
      +  \int^{\infty}_{-\infty}\; \frac{dq}{2\pi} 
           R(q|k)\; e^{iqx+i\alpha_0(q,\omega)z},
\end{eqnarray}
where the plane incident wave is given by:
\begin{eqnarray}
    \label{inc-field}
    \Phi_0(x,z|\omega) &=& \exp\left\{ikx-i\alpha_0(k,\omega)z\right\}
\end{eqnarray}
and $R(q|k)$ is the {\em scattering amplitude}. In the above
expressions, we have defined
\begin{eqnarray}
\alpha_0(q,\omega)&=&\sqrt{\left(\frac{\omega}{c}\right)^2-q^2}, \quad
 {\Re}\alpha_0(q,\omega)>0, {\Im}\alpha_0(q,\omega)>0.
\end{eqnarray}
Furthermore, the (longitudinal) momentum variables $q$ and $k$ are in
the radiative region related to respectively the scattering and
incident angle by 
\begin{mathletters}
  \label{mom-angel}
  \begin{eqnarray}
    k &=& \frac{\omega}{c}\sin\theta_0,\\
    q &=& \frac{\omega}{c}\sin\theta, 
  \end{eqnarray}
so that the $z$-components of the incident and scattering wavenumbers become 
  \begin{eqnarray}
    \alpha_0(k,\omega) &=& \frac{\omega}{c}\cos\theta_0, \\
    \alpha_0(q,\omega) &=& \frac{\omega}{c}\cos\theta.
  \end{eqnarray}
\end{mathletters}

The mean differential reflection coefficient~(DRC), also known as the
mean scattering cross section, is an experimentally accessible
quantity. It is defined as the fraction of the total, time-averaged,
incident energy flux scattered into the angular interval
$(\theta,\theta+d\theta)$. It can be shown to be related to the
scattering amplitude by the following expression~\cite{AnnPhys}:
\begin{eqnarray}
    \label{MDRC}
    \left< \frac{\partial R_s}{\partial \theta} \right>
    &=& \frac{1}{L}\;\frac{\omega}{2\pi c}\;\frac{\cos^2\theta}{\cos\theta_0}
      \left< \left| R(q|k)\right|^2\right>.    
\end{eqnarray}
Here $L$ denotes the length covered by the self-affine profile as
measured along the $x$-direction, and the other quantities have been
defined earlier.  The angle brackets denote an average over an
ensemble of realizations of the rough surface profiles.
Moreover, the momentum variables appearing in Eq.~(\ref{MDRC}) are
understood to be related to the angles $\theta_0$ and $\theta$
according to Eqs.~(\ref{mom-angel}).

We now impose the Kirchhoff approximation which  consists of locally replacing the surface by its
tangential plane at each point, and thereafter using the (local)
Fresnel reflection coefficient for the local angle of incidence to
obtain the scattered field.  Notice that dealing with a surface whose
scaling invariance range is bounded by a lower cut-off $\xi_{-}$ does
ensure that the tangential plane is well defined at every point.
Within the Kirchhoff approximation the scattering amplitude can be
expressed as~\cite{AnnPhys}:
\begin{mathletters}
  \label{R-amp}
  \begin{eqnarray}
    \label{scattering-amp}
    R(q|k) &=& \frac{-i}{2 \alpha_0(q,\omega)} 
    \int^{L/2}_{-L/2}dx \;
    e^{-iqx-i\alpha_0(q,\omega)\zeta(x)} {\cal N}_0(x|\omega),
  \end{eqnarray}
  where $ {\cal N}_0(x|\omega)$ is a source function defined by 
  \begin{eqnarray}
    {\cal N}_0(x|\omega) &=& 2\,\partial_n\!\!
    \left.\Phi_0(x,z|\omega)\right|_{z=\zeta(x)}.
  \end{eqnarray}
\end{mathletters}
Here $\partial_n$ denotes the (unnormalized) normal derivative defined
as $\partial_n=-\zeta'(x)\partial_x+\partial_z$.

By substituting the expression for the scattering amplitude ,
Eq.~(\ref{scattering-amp}), into Eq.~(\ref{MDRC}), one obtains an
expression for the mean differential reflection coefficient in terms
of the source function ${\cal N}_0(x|\omega)$; the normal derivative
of the total field evaluated on the rough surface.  After some
straightforward algebra where one takes advantage of the fact that the
self-affine surface profile function $\zeta(x)$ has stationary
increments, one obtains the following form for the mean differential
reflection coefficient
\begin{mathletters}
  \label{MDRC-mod}
  \begin{eqnarray}
    \left< \frac{\partial R_s}{\partial \theta} \right>
    &=&
    \D{   \frac{\omega}{2\pi c} \frac{1}{\cos\theta_0}
      \left( \frac{\cos \left[(\theta+\theta_0)/2\right]}
        {\cos \left[(\theta-\theta_0)/2\right]} \right)^2}
    \int^{L/2}_{-L/2}dv\; 
    \exp\left\{
      i\frac{\omega}{c}(\sin\theta-\sin\theta_0)v
    \right\} \Omega(v) ,
  \end{eqnarray}
  where 
  \begin{eqnarray}
    \label{Omega} 
    \Omega(v) &=& \left< \exp\left\{-i\frac{\omega}{c}[\cos\theta+\cos\theta_0]\,
        \Delta\zeta(v)\right\}\right>,
  \end{eqnarray}
\end{mathletters}
with $\Delta\zeta(v)=\zeta(x)-\zeta(x+v)$.  Note that the statistical
properties of the profile function, $\zeta(x)$, enters
Eqs.~(\ref{MDRC-mod}) only through $\Omega(v)$.  With the height
distribution $p(\delta \zeta; \delta x)$ introduced earlier,
Eq.~(\ref{dist}), one may now analytically calculate the ensemble
average contained in $\Omega(v)$.  For a Gaussian self-affine surface
one gets
\begin{eqnarray}
    \label{Omega-2} 
    \Omega(v) &=& 
      \int^{\infty}_{-\infty} dz\;
      \exp\left\{ -i\frac{\omega}{c}\left(
      \cos\theta+\cos\theta_0\right)z\right\}
          p(z; v)
      \nonumber \\
      &=&  
      \exp\left\{ -\left(\frac{\omega}{c} 
                     \frac{\cos\theta+\cos\theta_0}{\sqrt{2}}
                       s(\lambda)\lambda^{1-H}v^H
                   \right)^2
             \right\}.
\end{eqnarray}
By in Eq.~(\ref{MDRC-mod}) making the change of variable
\begin{eqnarray}
    \label{s-var} 
u &=& v \left[ \frac{\omega}{c} 
                       \frac{\cos\theta+\cos\theta_0}{\sqrt{2}} s(\lambda)\lambda^{1-H}
        \right]^{1/H} \;,
\end{eqnarray}
and letting the length of the profile extend
to infinity, $L\rightarrow \infty$, one finally obtains the following
expression for the mean differential reflection coefficient:
\begin{mathletters}
    \label{final-result}
\begin{eqnarray}
    \label{MDRC-final}
    \left< \frac{\partial R_s}{\partial \theta} \right>
    &=&
    \frac{s(\lambda)^{-\frac{1}{H}}
a^{-(\frac{1}{H}-1)}}{\sqrt{2}\, \cos\theta_0 }
     \frac{
    \cos\frac{\theta+\theta_0}{2}}{\cos^3\frac{\theta-\theta_0}{2}}  
 {\cal L}_{2H}\left(\frac{ \sqrt{2}\tan\frac{\theta-\theta_0}{2} }{ a^{\frac{1}{H}-1}
s(\lambda)^{\frac{1}{H}} }\right),  
\end{eqnarray}
where 
\begin{eqnarray}
  \label{u-def}
  a &=& 2\pi\sqrt{2}
  \cos\frac{\theta+\theta_0}{2}\cos\frac{\theta-\theta_0}{2},
\end{eqnarray}
and ($0<\alpha\leq 2$)
\begin{eqnarray}
  \label{Levy-distribution-1}
  {\cal L}_\alpha(x) &=& \frac{1}{2\pi} \int^\infty_{-\infty} dk \; 
  e^{ikx} e^{-\left| k \right|^\alpha}. 
\end{eqnarray}
\end{mathletters}
The quantity ${\cal L}_\alpha(x)$ is known as the centered symmetric
L\'evy stable distribution of index (or order) $\alpha$~\cite{Levy}.
This distribution can only be expressed in closed form for some
particular values of $\alpha=$; $\alpha=1$ and $\alpha=2$correspond to
the Cauchy-Lorentzian and Gaussian distributions respectively, ${\cal
L}_{1/2}$ and ${\cal L}_{1/3}$ can be expressed from special
functions. When the $\alpha$-index in the L\'evy distribution ${\cal
L}_\alpha(x)$ is lowered from its upper value $\alpha=2$ (Gaussian
distribution), the resulting distribution develops a sharper peak at
$x=0$ while at the same time its tails become fatter.  It is
interesting to note from Eqs.~(\ref{final-result}) that the
wavelength, $\lambda=2\pi c/\omega$, only comes into play through the
slope $s(\lambda)$. The behavior of the scattered intensity is thus
entirely determined by this typical slope $s(\lambda)$ and the
roughness exponent $H$.


\begin{figure}[t!]
  \begin{center}
    \epsfig{file=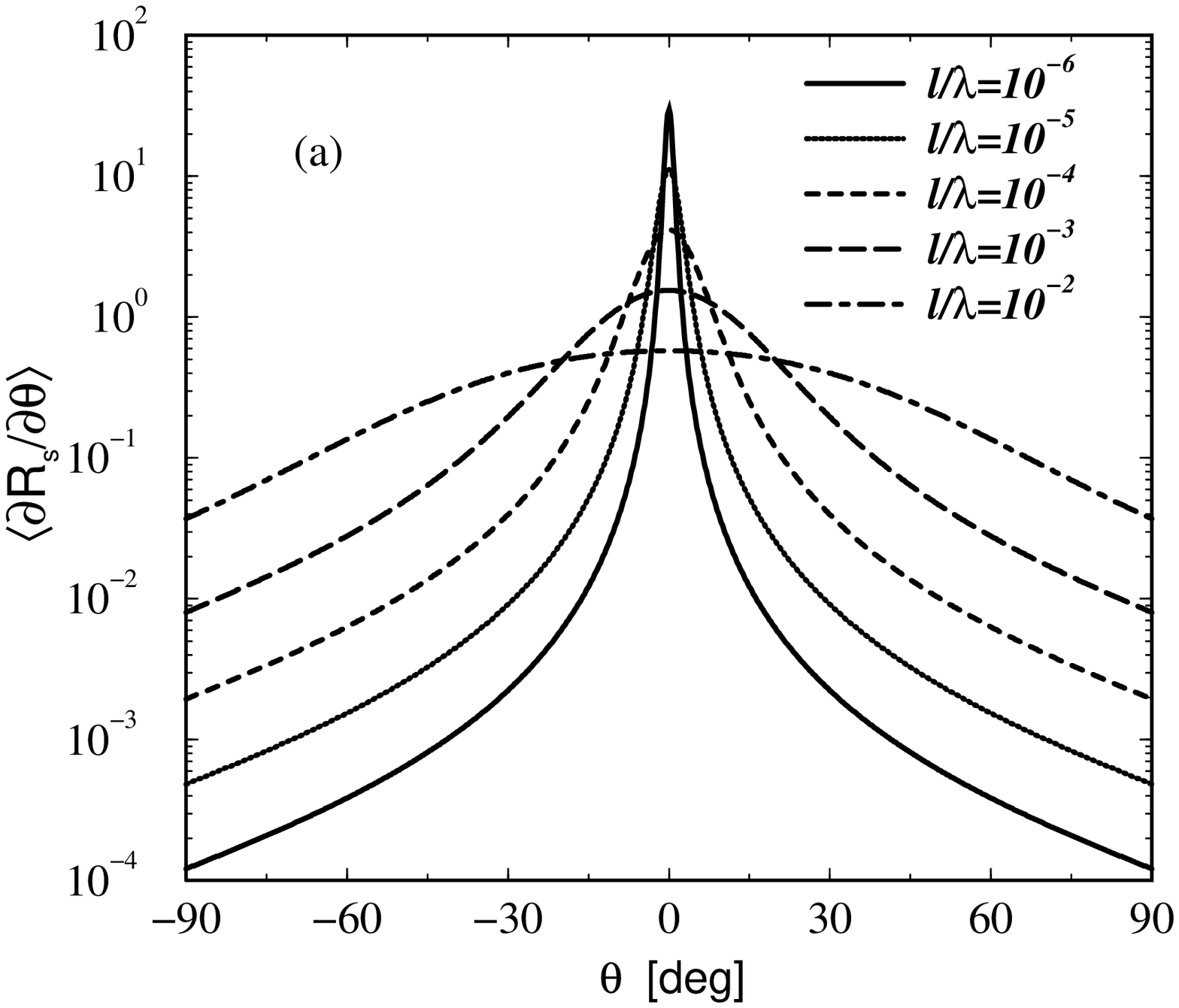,width=7.5cm,height=5.5cm}
    \qquad
    \epsfig{file=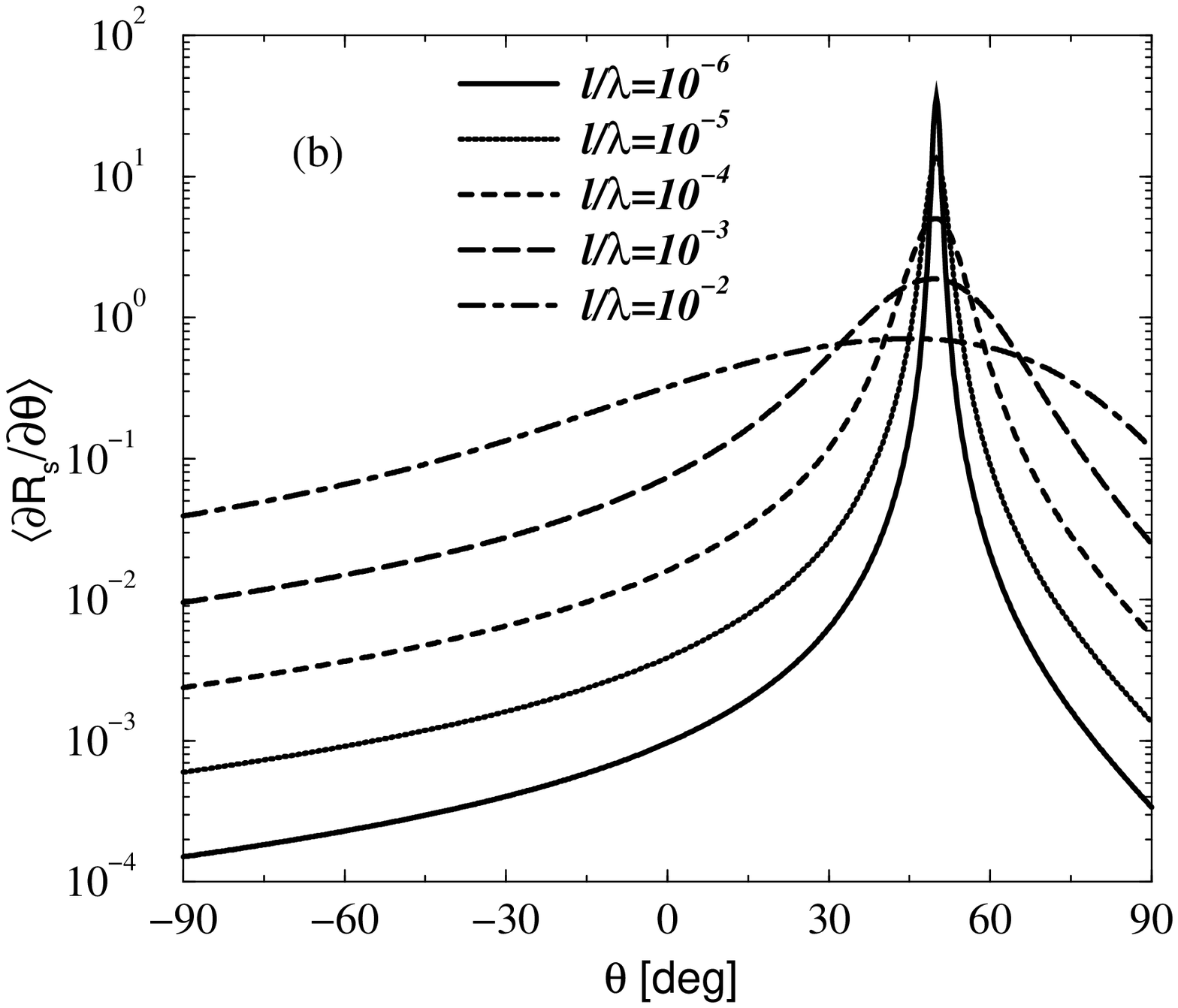,width=7.5cm,height=5.5cm}
  \end{center}
  \caption{The mean differential reflection coefficient,
    $\left<\partial R_s/\partial \theta\right>$, vs. scattering angle,
    $\theta$, for a perfectly conducting self-affine surface.  The
    plotted curves are the prediction of
    Eqs.~(\protect\ref{final-result}). The Hurst exponent in all cases
    are $H=0.7$ and topothesies $\ell$ are ranging from
    $\ell=10^{-2}\lambda$ ($S(\lambda=0.016$) down to $10^{-6}\lambda$
    ($s(\lambda)=0.25$) as indicated in the figures. The angles of
    incidence were (a) $\theta_0=0^\circ$ and (b) $\theta_0=0^\circ$.}
    \label{Fig:Analytic-results}
\end{figure}


In Figs.~\ref{Fig:Analytic-results} we show the mean differential
reflection coefficient as obtained from Eqs.~(\ref{final-result}) for
Hurst exponent $H=0.7$ and different values of the slope $s(\lambda)$
ranging from $0.016$ to $0.25$. The angles of incidence were $\theta
_0=0^\circ$~(Fig.~\ref{Fig:Analytic-results}a) and
$50^\circ$~(Fig.~\ref{Fig:Analytic-results}b). It is observed from
these figures that as the amplitude parameter $s(\lambda)$ is
decreased, while keeping the other parameters fixed, the portion of
the scattered intensity scattered diffusely is reduced, while the
power-law behavior found for the non-specular directions survives
independently (within single scattering) of the amount of light
scattered specularly.  Furthermore, as the Hurst exponent is
decreased~(results not shown), and thereby making the topography 
rougher at small scale, the mean DRC gets a larger contribution from
diffusely scattered light. This is a direct consequence of the
properties of the L\'evy distribution mentioned above.

In order to discuss the features of the mean DRC which can be seen in
Figs.~\ref{Fig:Analytic-results} we will now proceed by discussing the
behavior of the specular and diffuse contribution to $\left< \partial
  R_s/\partial \theta \right>$, {\it i.e.} close and far away from the
scattering angle $\theta=\theta_0$ respectively.

\subsection{The specular contribution}

We start by considering the specular contribution to the mean
differential reflection coefficient.  This is done by taking advantage
of the asymptotic expansion of the L\'evy distribution around
zero~\cite{Levyexp}
\begin{eqnarray}
  \label{Levyzero} 
  {\cal L}_{\alpha}(x) &=& \frac{1}{\pi\alpha} \Gamma \left( \frac{1}{\alpha} \right)
  \left[ 1 - \frac{\Gamma \left( \frac{3}{\alpha} \right)}{2\Gamma
      \left( \frac{1}{\alpha} \right) } x^2 \right]+{\cal O}(x^4).
\end{eqnarray}
By substituting this expression into Eqs.~(\ref{final-result}) one
finds that the mean DRC around the speular direction
$\theta=\theta_0$ should behave as follows ($\delta\theta \ll 1$)
\begin{eqnarray}
  \label{specular-behavior}
\left. \left< \frac{\partial R_s}{\partial \theta}\right>\right|_{\theta=\theta_0+\delta\theta}
    &=&
\frac{\Gamma \left( \frac{1}{2H} \right)}{
2\sqrt{2}\pi H \left(2\sqrt{2}\pi \cos\theta_0 \right)^{\frac{1}{H}-1}
s(\lambda)^{1/H} }
  \nonumber \\
  & &\quad \times
\left[
 1+\delta\theta \frac{1-2H}{2H}\tan\theta_0
 + \frac{(\delta\theta)^2}{4}
\left(1+\frac{1+H}{2H^2}\tan^2\theta_0 
-\frac{\Gamma(\frac{3}{2H})}{\Gamma(\frac{1}{2H})
\left( 2\sqrt{2} \pi  \cos\theta_0 
    \right)^{\frac{2}{H}-2} s(\lambda)^{2/H}}  \right)
\right].
\end{eqnarray}
From this expression it follows that the amplitude of the specular
peak should scale as
\begin{eqnarray}
    \label{amplitude}
\left. \left< \frac{\partial R_s}{\partial \theta} \right>
\right|_{\theta=\theta_0}
\simeq 
\frac{\Gamma(\frac{1}{2H}) }
   {2\sqrt{2}\pi H \left(2\sqrt{2}\pi \cos\theta_0 
   \right)^{\frac{1}{H}-1}s(\lambda)^{1/H} },
\end{eqnarray}
and that the peaks half width at half maximum, $w$, should be given by

\begin{figure}[t!]
  \begin{center}
    \epsfig{file=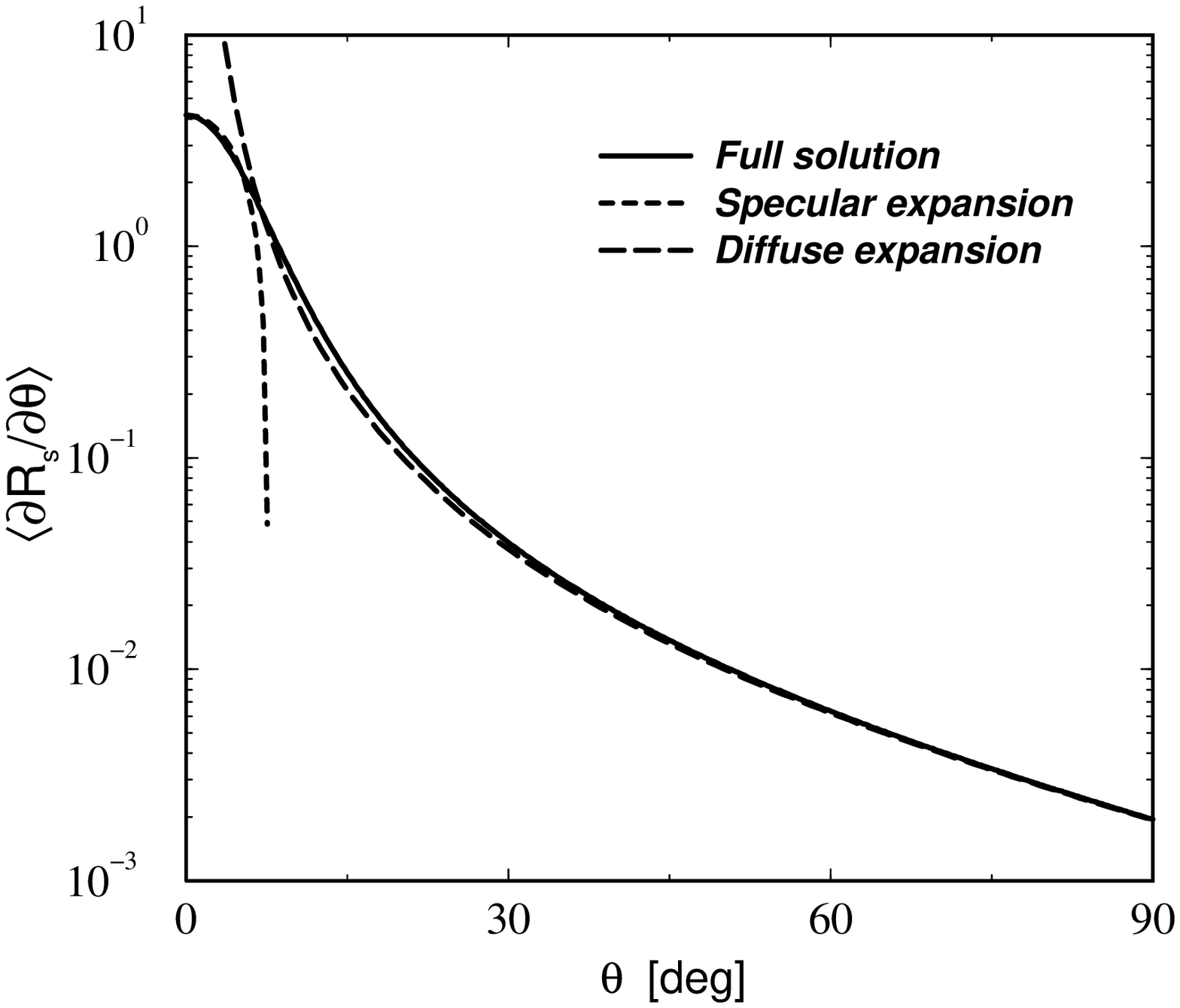,width=7.5cm,height=5.5cm}
  \end{center}
  \caption{The full single scattering solution (solid line),
    Eqs.~(\protect\ref{final-result}), for the mean differential
    reflection coefficient vs. scattering angle $\theta$ for a
    perfectly conducting self-affine surface compared to its
    specular (dotted line) and diffuse expansions (dashed line) as
    given by respectively Eqs.~(\protect\ref{specular-behavior}) and
    (\protect\ref{diffuse-behavior}). The surface parameters used
    were $H=0.7$ and $\ell=10^{-4}\lambda$ ($s(\lambda)=0.063$),
    and the light was incident normally onto the rough surface. }
  \label{Fig:spec-diff-expansion}
\end{figure}


\begin{eqnarray}
  \label{exactwidth}
w\left(H,s(\lambda),\theta_0\right)  &\simeq&  2 \sqrt{ 
     \frac{\Gamma(\frac{1}{2H})}{\Gamma(\frac{3}{2H})} }
   \left( 
        2\sqrt{2}\pi \cos\theta_0 
   \right)^{\frac{1}{H}-1 }s(\lambda)^{1/H} \;.
\end{eqnarray}

It is worth noting that in the above expression the width of the
specular peak depends on the wavelength $\lambda$ {\it via} the
typical slope over one wavelength $s(\lambda)$. In case of Gaussian
correlations, there would have been no dependence on the wavelength,
the peak width $w$ being simply proportional to the ratio
$\sigma/\tau$, RMS roughness over correlation length.

In order to test the quality of the specular expansion,
Eq.~(\ref{specular-behavior}), we show in
Fig.~\ref{Fig:spec-diff-expansion} a comparison of this expression
with the full single scattering solution obtained from
Eqs.~(\ref{final-result}) for a surface of roughness exponent $H=0.7$
and of slope over the wavelength $s(\lambda)=0.063$
($\ell=10^{-4}\lambda$) in case of normal incidence. The amplitude of
the specular peak is seen to be nicely reproduced, but this expansion
is only valid within a rather small angular interval around the
specular direction $\theta=\theta_0$.

It is interesting to notice that in the case of a non-zero angle of
incidence, $\theta_0\neq 0^\circ$, the specular peak is slightly
shifted away from its expected position $\theta=\theta_0$ due to the
presence of a non-vanishing term in Eq.~(\ref{specular-behavior})
linear in $\delta\theta$.  In this case the apparent specular peak is
located at $\theta=\theta_0 + \Delta\theta_0$, where
$\Delta\theta_0$~($\Delta\theta_0 \sim w^2\ll w$) scales as
\begin{eqnarray}
\Delta\theta_0 \simeq 
 \frac{2H-1}{H} \frac{\Gamma(\frac{1}{2H})}{\Gamma(\frac{3}{2H})}
   \tan\theta_0 \left( 2\sqrt{2}\pi\cos\theta_0 
                 \right)^{\frac{2}{H}-2} s(\lambda)^{2/H}
   \;=\; \frac{2H-1}{4H}\tan\theta_0  w^2(H,s(\lambda),\theta_0) 
\end{eqnarray}
Such a shift has not, to our knowledge, been reported earlier for non
self-affine (or non fractal) surfaces. Hence, due to the self-affinity
of the random surface, we predict a shift, $\Delta\theta_0$, in the
specular direction as compared to its expected position at
$\theta=\theta_0$.  Notice that this shift vanishes for a Brownian
random surface ($H=1/2$).  Moreover, for a persistent surfaces profile
function ($H>1/2$) the shift is positive while it becomes negative for
anti-persistent profile ($H<1/2$).  Unfortunately the specular shift
$\Delta\theta_0$ is probably too small to be observable experimentally
for realizable self-affine parameters.


\subsection{The diffuse component}

We now focus on the diffuse component to the mean differential
reflection coefficient, {\it i.e.} the region where $\theta$ is far away
from $\theta=\theta_0$. Now, using the expansion of the L\'evy
distribution at infinity (the Wintner development)~\cite{Levyexp}
\begin{eqnarray}
  \label{Levyinf} 
  {\cal L}_{\alpha}(x) &=&
  \frac{\Gamma (1+\alpha)}{\pi |x|^{1+\alpha}} \sin \left(
    \frac{\alpha\pi}{2} \right)+ {\cal O} \left(
    \frac{1}{|x|^{1+2\alpha}} \right),
\end{eqnarray}  
we get the following expression for the diffuse component of the mean
DRC ($\theta\neq \theta_0$)
\begin{eqnarray}
  \label{diffuse-behavior}
\left< \frac{\partial R_s}{\partial \theta}\right> \simeq
   \frac{\Gamma(1+2H) \sin(\pi H)} {\left(4\pi\right)^{2H-1}}
\frac{s(\lambda)^2}{\cos\theta_0 }
\frac{\left|\cos\frac{\theta+\theta_0}{2}\right|^{3-2H}}   
     {  \left| \sin\frac{\theta-\theta_0}{2}\right|^{1+2H}} \;.
\end{eqnarray}

In Fig.~\ref{Fig:spec-diff-expansion} the above expression is compared
to the prediction of Eqs.~(\ref{final-result}).  We observe an
excellent agreement for angular distances larger than ten degrees.
Moreover, it should be noticed from Eq.~(\ref{diffuse-behavior}) that
the mean DRC is predicted to decay as a power-law of exponent $-1-2H$
as we move away from the specular direction.  For smooth surfaces
(corresponding to small values of $s(\lambda)$) this behavior results
directly from a perturbation approach where the scattered intensity
derived directly from the power density function of the surface. As
shown above, in the case of self-affine surfaces the latter is a power law
of exponent $-1-2H$. Our results extend then the validity of this
power law regime to steeper surfaces.

\section{Scaling analysis}

It is interesting that most of the non-trivial scaling results derived
above can be retrieved via simple dimensional arguments. Let us
examine the intensity scattered in direction $\theta$; in a naive
Huyghens framework two different effects will compete to destroy the
coherence of two source points on the surface {\it i}) the angular
difference separating $\theta$ from the specular direction and {\it
ii}) the roughness. Considering two points separated by a horizontal
distance $\Delta x$ and a vertical distance $\Delta z$, we can define
the retardation due to these two effects:
$$
\Delta c_{ang} = (\sin \theta -\sin \theta_0) \Delta x \;, \quad
\Delta c_{rough}=(\cos \theta +\cos \theta_0) \Delta z\;.
$$
This allows us to define two characteristic (horizontal) lengths
$\delta_{ang}$ and $\delta_{rough}$ of the scattering system
corresponding to the distances between two points of the surface such
that $\Delta c_{ang}$ and $\Delta c_{rough}$ are equal to the
wavelength $\lambda$. Taking into account the self-affine character of
the surface, we get:
$$
\delta_{ang}=\frac{\lambda}{\sin \theta -\sin \theta_0} \;, \quad
\delta_{rough}=\frac{\lambda}{(\cos \theta +\cos \theta_0)^{1/H}}s(\lambda)^{-1/H}\;.
$$
The coherence length on the surface depends on the relative magnitude  of these two
characteristic lengths. For scattering angles close to the specular
direction, we have $\delta_{rough} \ll \delta_{ang}$ and  for large scattering
angles $\delta_{ang} \le \delta_{rough}$ and the diffuse tail is
controlled by the angular distance to the specular direction. In
general we can evaluate the competition of these two effects and their
consequences on the scattering cross-section by the simple ratio of
the two characteristic lengths:
$$
\chi=\frac{ \delta_{rough}}{\delta_{ang}}
=\frac{\sin \theta -\sin \theta_0}{(\cos \theta +\cos \theta_0)^{1/H}}
s(\lambda)^{-1/H}\;.
$$
We can then describe our scattering system with this unique variable
$\chi$ which takes into account the incidence and scattering
directions, the roughness parameters of the surface and the
wavelength. A direct application is the determination of the angular
width $w$ of the specular peak. The transition between the specular peak
and the diffuse tail is simply defined by $\chi=1$ which leads to:
$$
w\simeq [2s(\lambda)]^{1/H} (\cos \theta_0)^{1/H-1}
$$
which is identical to the exact result (\ref{exactwidth}) apart from a
numerical constant. Assuming that most of the intensity is scattered
within the specular peak, we obtain {\it via} the energy conservation 

$$
\left.\left< \frac{\partial R_s}{\partial \theta}\right>\right|_{\theta=\theta_0}
\simeq \frac{1}{w} \simeq [2s(\lambda)]^{-1/H} (\cos \theta_0)^{1-1/H}
$$

Forgetting the numerical constants, we can thus rewrite the scattering cross-section as
$$
\left< \frac{\partial R_s}{\partial \theta}\right>
=\frac{(\cos \theta_0)^{1-1/H}}{s(\lambda)^{1/H}}
\Psi(\chi)
$$
When approaching the specular direction we note that $\delta_{ang}$
diverges whereas $\delta_{rough}$ saturates at a finite value
independent on the angular direction. In this specular direction, the
scattering process is thus controlled by only the latter length and
does not depend on the ratio $\chi=\delta_{rough}/\delta_{ang}$. This imposes:
$$
\Psi(\chi)\simeq 1,\qquad (\chi \ll 1).
$$
The argument $\chi$ being inversely proportional to the quantity
$s(\lambda)^{1/H}$ which is nothing but a roughness amplitude
parameter, the behavior of $\Psi$ for large arguments can be found by
matching our expresion with the limit of very smooth surfaces.  In
this limit a simple perturbation approach leads to:
$$
\left< \frac{\partial R_s}{\partial \theta}\right>
\propto {\cal P}\left[\frac{2\pi}{\lambda}(\sin \theta -\sin \theta_0)\right],
$$
where ${\cal P}$ is the power density function of the height
profile. In the case of a self-affine profile of roughness exponent
$H$, we have ${\cal P}(k) \propto k^{-1-2H}$. One can check that this
 can only be consistent with the same power law behavior for $\Psi$:

$$
\Psi(\chi)\propto \chi^{-1-2H},  \qquad (\chi \gg 1).
$$

\section{Numerical simulation results and discussion}

The results obtained in the previous sections were all based on the
Kirchhoff approximation, and will therefore only be accurate in cases
where single scattering is dominating. In this section, however, we
will therefore no longer restrict ourselves to single scattering, but
instead include any higher order scattering process.  This is
accomplished by a rigorous numerical simulations approach which will
be described below. This approach will also serve as an independent
check of the correctness of the analytic results~(\ref{final-result}),
and the results that can be derived thereof. Furthermore, it will also
provide valuable insight into which part of parameter space is
dominated by single scattering processes, and thus where
formulae~(\ref{final-result}) can be used with confidence.

The rigorous numerical simulation calculations for the mean
differential reflection coefficient were performed for a plane
incident $s$-polarized wave  scattered from a perfectly
conducting rough self-affine surface.  Such simulations were done by
the now quite standard extinction theorem technique~\cite{AnnPhys}.
This technique amounts to using Green's second integral identity to
write down the following inhomogeneous Fredholm equation of the second
kind for the source function ${\cal N}(x|\omega)$~(see
Refs.~\cite{Thorsos1,Thorsos2}):
\begin{mathletters}
  \begin{eqnarray}
    \label{int-eq}
    {\cal N}(x|\omega) =  
    2 {\cal N}_0(x|\omega) - 2 \, {\cal P}\!\! \int dx'\;  
       \left. \partial_n G_0(x,z|x',z')\right|_{z'=\zeta(x')} 
         {\cal N}(x'|\omega).
  \end{eqnarray}
  In this equation 
  \begin{eqnarray}
    {\cal N}(x|\omega) &=& \left. \partial_n 
            \Phi(x,z|\omega)\right|_{z=\zeta(x)},
  \end{eqnarray}
  where $\partial_n = -\zeta'(x)\partial_x+\partial_z$ is the
  (unnormalized) normal derivative of the {\em total} electric field
  $\Phi=E_y$ evaluated on the randomly rough self-affine surface,
  ${\cal N}_0(x|\omega)$ has been defined earlier as the normal
  derivative of the incident field, and ${\cal P}$ is used to denote
  the principle part of the integral. Moreover, $G_0(x,z|x',z')$ is
  the (two-dimensional) free space Greens function defined by
  \begin{eqnarray}
    \label{Greens-function}
    G_0(x,z|x',z')  &=& i\pi
    H^{(1)}_{0}\left(\frac{\omega}{c}\left|{\bf r}-{\bf r}'\right|\right),
  \end{eqnarray}
  where ${\bf r}=(x,z)$, ${\bf r'}=(x',z')$ and $H^{(1)}_{0}(x)$ denotes the 0 th order Hankel
  function of the first kind~\cite{Stegun}.
\end{mathletters}
By taking advantage of Eq.~(\ref{scattering-amp}) which relates the
scattering amplitude to the normal derivative of the total field on
the random surface, the scattering amplitude can easily be calculated,
and from there the mean differential reflection coefficient.  It
should be noticed  that the Kirchhoff approximation used in the previous
section to obtain the analytical results~(\ref{final-result}), is
obtained from by Eq.~(\ref{int-eq}) by neglecting the last (integral)
term that represents multiple scattering.  By using a numerical
quadrature scheme~\cite{NR}, the integral equation,
Eq.~(\ref{int-eq}), can be solved for any given realization of the
surface profile $\zeta(x)$.  From the knowledge of ${\cal
  N}(x|\omega)$ one might then easily calculate the mean DRC.

Randomly rough Gaussian self-affine surfaces of given Hurst exponent
were generated by the Fourier filtering method~\cite{Feder} (see
Eq.~(\ref{SA-power-spec})), {\it i.e.}  in Fourier space to filter
complex Gaussian random uncorrelated numbers by a decaying power-law
filter of exponent $-H-1/2$ and thereafter transforming this sequence
into real space.  The topothesis (or slope) of the surfaces were then
adjusted to the desired values, $\ell$, by taking advantage of
Eq.~(\ref{Eq:sigma}). This is done by first calculating the topothesy,
$\ell_0$, of the original surface over its total length and thereafter
rescaling the profile by $(\ell_0/\ell)^{1-H}$, where $\ell$ is the
desired topothesy. In order to having enough statistical information
to be able to calculate a well-defined topothesy $\ell_0$, we in fact
used a window size slightly smaller than the total length of the
surface.

\begin{figure}[t!]
  \begin{center}
         \epsfig{file=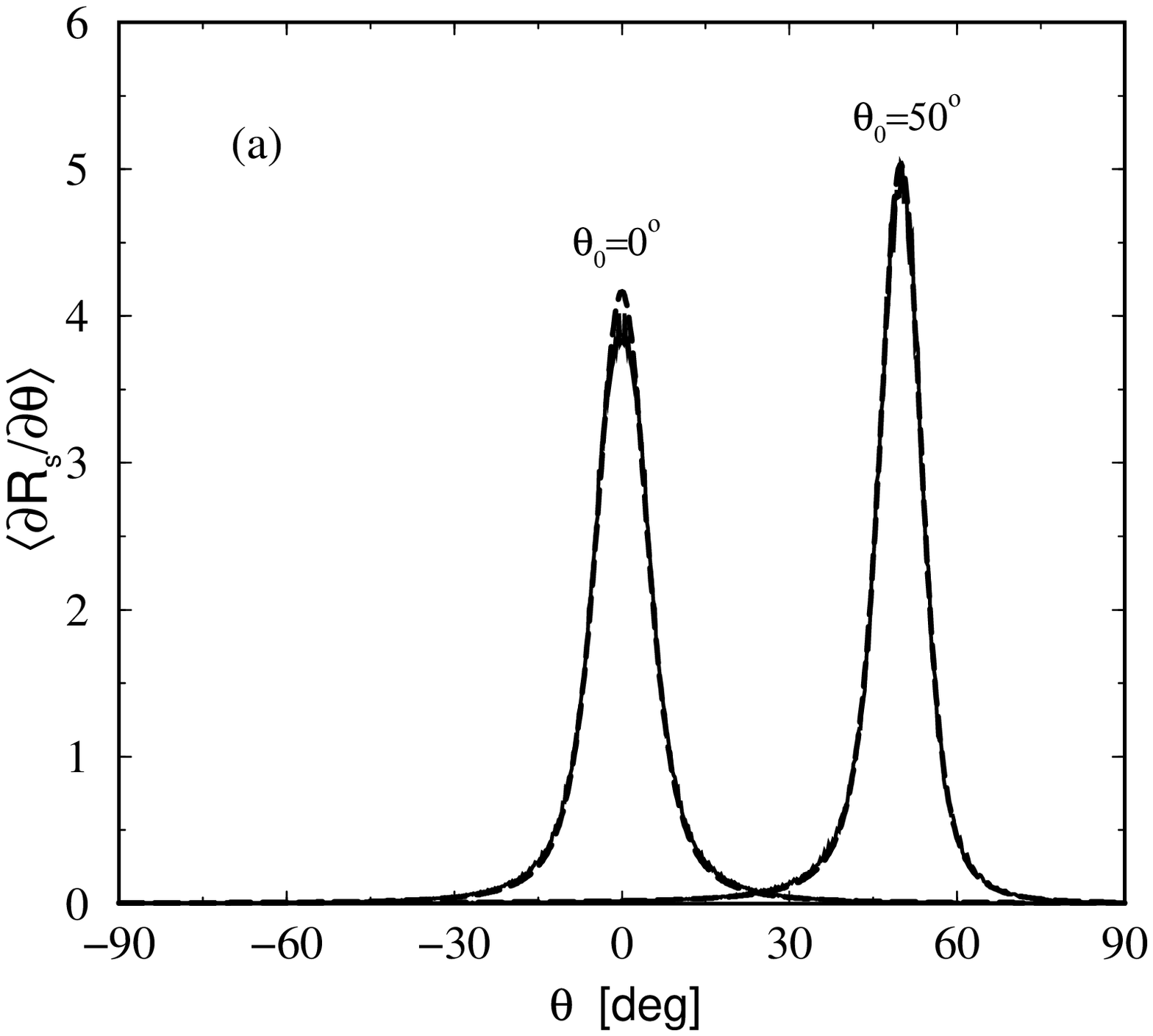,width=7.5cm,height=5.5cm}
         \qquad
         \epsfig{file=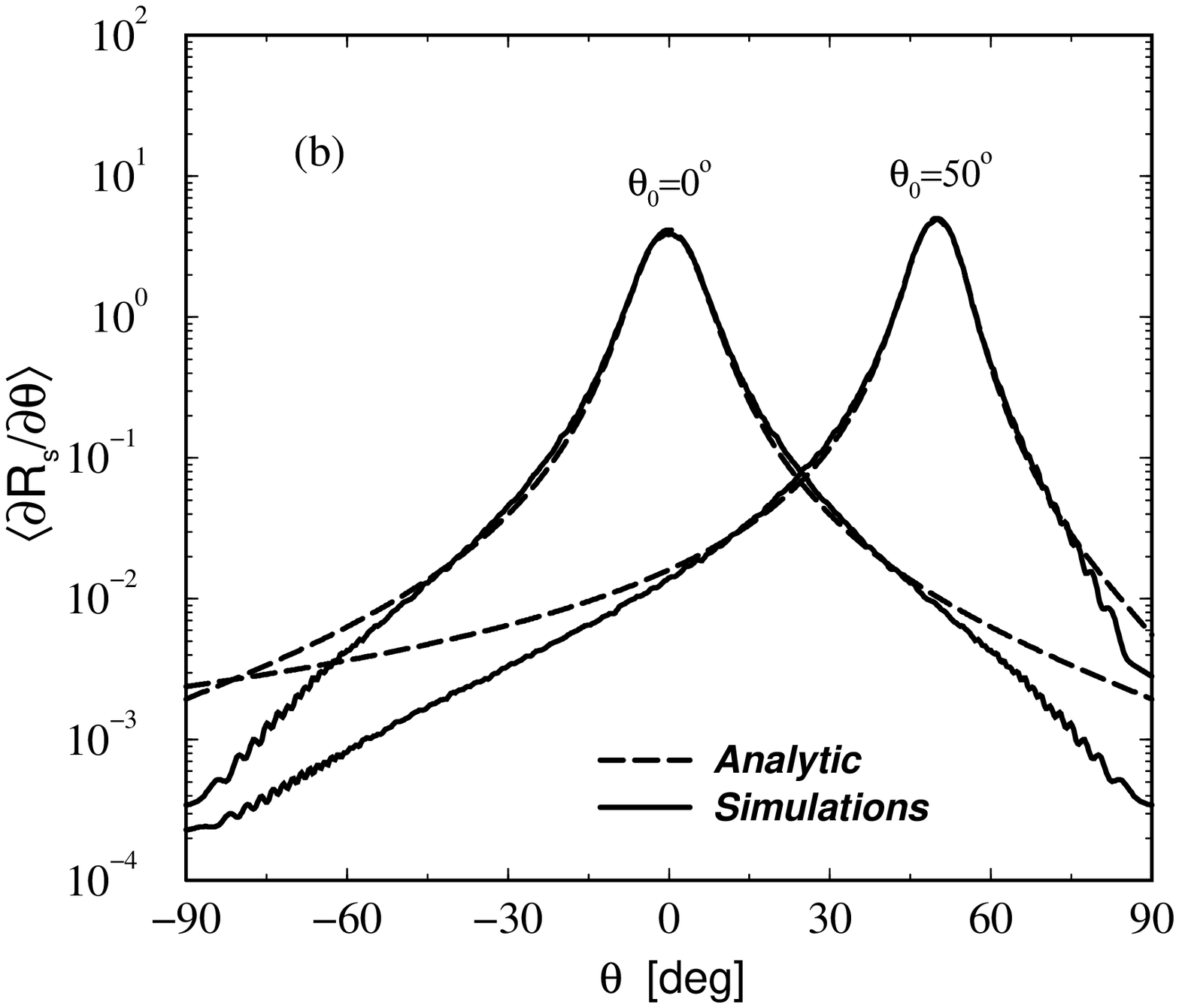,width=7.5cm,height=5.5cm}
     \end{center}
    \caption{A comparison plotted in  (a) linear- and (b) linear-log scale, 
      of the mean differential reflection coefficient $\left<\partial
        R_s/\partial \theta\right>$ vs. scattering angle $\theta$ for
      a perfectly conducting self-affine surface obtained by a
      rigorous numerical simulations approach (solid lines), and
      therefore including all possible multiple-scattering processes,
      and the single-scattering results obtained from
      Eqs.~(\protect\ref{final-result})~(dashed lines). The surface
      parameters were $H=0.7$ and $\ell=10^{-4}\lambda$
      ($s(\lambda)=0.063$) with $\lambda=612.7{\rm nm}$. The angles of
      incidence for the light were $0^\circ$ and $50^\circ$ as
      indicated in the figure.}
    \label{Fig:Compare}
\end{figure}


By the methods just described, we have performed rigorous numerical
simulations for the mean differential reflection coefficient,
$\left<\partial R_s/\partial \theta\right>$, in the case of a
$s$-polarized plane incident wave of wavelength $\lambda=2\pi
c/\omega=612.7~{\rm nm}$ that is scattered from a perfectly conducting
self-affine surface characterized by the Hurst exponent, $H$, and the
topothesy $\ell$. For all simulation results shown, the length of the
surface was $L=100\lambda$, and the spatial discretization length was
$\Delta x\simeq \lambda/10$. All simulation results presented were
averaged over $N_\zeta=1000$ surface realizations (or more).
Furthermore, in order to check the quality of the numerical
simulations, both reciprocity and unitarity were checked for all
simulation results. It was found for all cases considered that the
reciprocity was satisfied within the noise level of the calculations,
while the unitarity was fulfilled within an error of a fraction of a
percent.

In Figs.~\ref{Fig:Compare} the mean differential reflection
coefficients for a surfaces characterized by the parameters $H=0.7$
and $\ell=10^{-4}\lambda$ ($s(\lambda)=0.063$) are presented. The
angles of incidence of the light were $\theta_0=0^\circ$ and
$50^\circ$ as indicated in the figures.  The solid lines represents
the numerical (multiple-scattering) simulation results while the
dashed lines are the (single-scattering) prediction of
Eqs.~(\ref{final-result}).  As can be seen from
Fig.~\ref{Fig:Compare}a the correspondence is quite good between the
analytic results and those obtained from the numerical simulations.
To allow a better comparison for large scattering angles we present in
Fig.~\ref{Fig:Compare}b the results of Fig.~\ref{Fig:Compare}a, but
now in a linear-log scale.  From this figure it is apparent that for
the largest scattering angles there are some disagreements between the
analytic and numerical results.  The analytic results tend to
overestimate the mean DRC in these regions.  This discrepancy stems
from the fact that multiple scattering is not included in the
analytical results.  Part of the light that according to single
scattering would have been scattered into large scattering angles are
now due to multiple scattering processes, scattered back into smaller

\vspace*{0.5cm}
\begin{figure}[t!]
  \begin{center}
    \epsfig{file=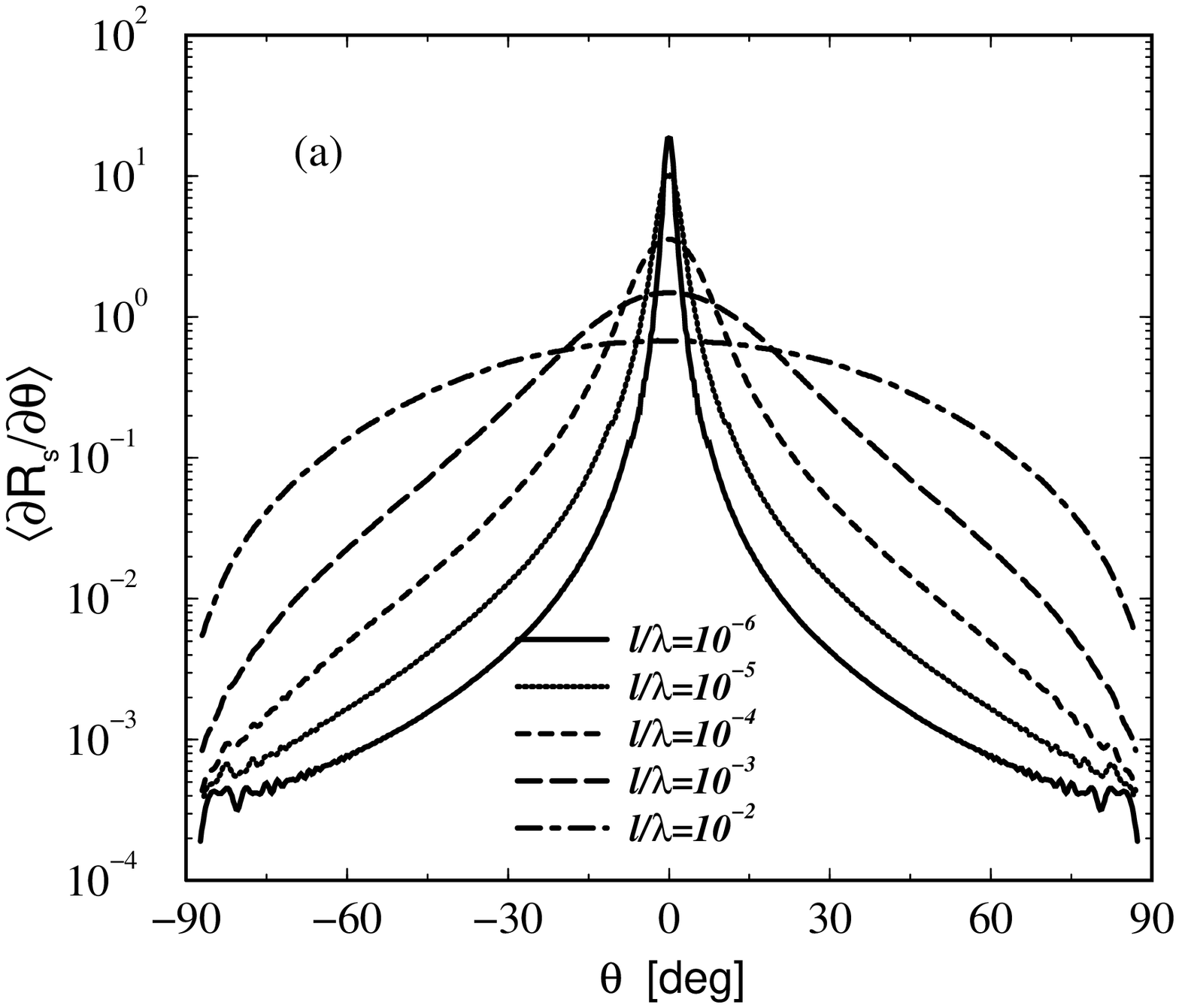,width=7.5cm,height=5.5cm}
    \qquad 
    \epsfig{file=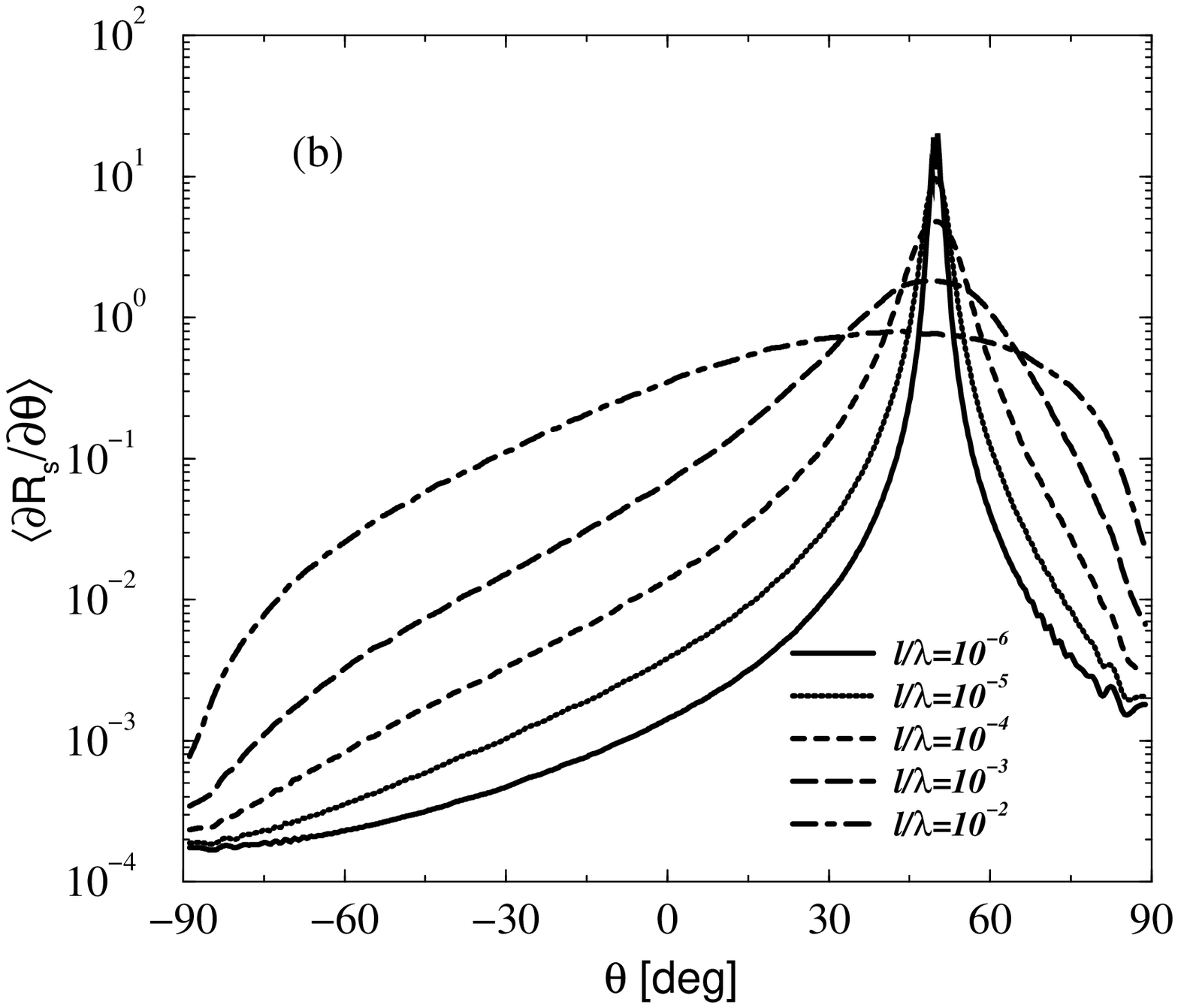,width=7.5cm,height=5.5cm}
  \end{center}
  \caption{The same as Figs.~\protect\ref{Fig:Analytic-results}
    (single scattering results), but now using a rigorous numerical
    simulation approach (see text for details) that incorporate all
    higher order scattering processes.}
  \label{Fig:Several-DRC}
\end{figure}


\begin{figure}[t!]
  \begin{center}
    \epsfig{file=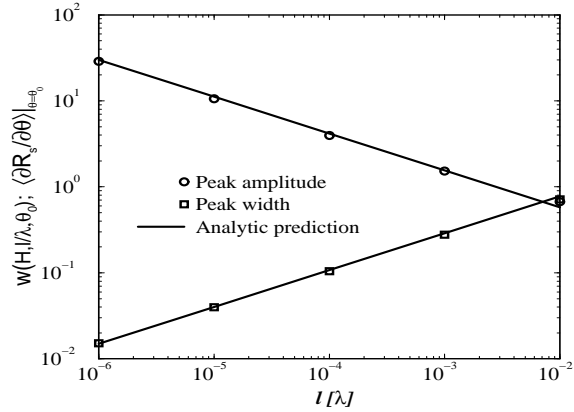,width=7.5cm,height=5.5cm}
  \end{center}
  \caption{The specular peak amplitude, 
    $\left. \left< \partial R_s/\partial \theta \right>
    \right|_{\theta=\theta_0}$, and its half width at half maximum,
    $w(H,\ell/\lambda,\theta_0)$ as a function of the topothesy
    $\ell$. The angle of incidence was in both cases
    $\theta_0=0^\circ$. The solid lines are analytical results
    obtained from Eqs.~(\protect\ref{amplitude}) and
    (\protect\ref{exactwidth}) respectively, while the circles
    (amplitudes) and the squares (widths) are obtained from the
    numerical simulations results shown in
    Fig.~\protect\ref{Fig:Several-DRC}a. }
  \label{Fig:Amplitude-Width}
\end{figure}

\noindent 
angles. This results in {\em smaller} values for $\left<\partial
  R_s/\partial \theta\right>$ for the largest scattering angles. Since
the unitarity condition, $\int^{\pi/2}_{-\pi/2} \left<\partial
  R_s/\partial \theta\right> \, d\theta =1$, is satisfied for a
perfectly reflecting surface, this large angle reduction of the mean
DRC has to be compensated by an increase for other scattering angles.
In the case of normal incidence ($\theta_0=0^\circ$) say, this
increase can be seen in the region around $|\theta|\sim 25^\circ$
where the numerical simulation results are bigger then the
corresponding single-scattering analytical results.  The same behavior
can be observed for an angle of incidence of $50^\circ$.

We give in Figs.~\ref{Fig:Several-DRC} the numerical simulation
results for five different values of the topothesy ranging from
$\ell=10^{-6}\lambda$ ($S(\lambda=0.016$) up to $10^{-2}\lambda$
($s(\lambda)=0.25$). These multiple-scattering results should be
compared to the results of Figs.~\ref{Fig:Analytic-results} which show
the corresponding curves obtained from Eqs.~(\ref{final-result}).  The
roughness exponent used in the simulations leading to the results of
Figs.~\ref{Fig:Several-DRC} was in all cases $H=0.7$, while for the
angles of incidence we used $\theta_0=0^\circ$
(Fig.~\ref{Fig:Several-DRC}a) and $\theta_0=50^\circ$
(Fig.~\ref{Fig:Several-DRC}b).  The height standard deviation as
measured over the whole length of the surface, $L=100\lambda$, was
according to Eq.~(\ref{Eq:sigma}) ranging from $\sigma(L)=0.4\lambda$
for the smallest topothesy up to as large as $6.3\lambda$ for the
largest.  The fact that we did not really use the total length, $L$,
during the surface generation when adjusting the topothesy, but
instead a slightly smaller fraction of this length, did not seem to
affect the height standard deviation to a large degree. In fact it was
found numerically that the {\sc RMS}-heights of the generated surfaces
were only a few percent lower then the one obtained from using
Eq.~(\ref{Eq:sigma}) and we will therefore in the following use this
equation in estimating the {\sc RMS}-height of the surfaces.
According to optical criterion these surface roughness correspond to
rather rough surfaces.  In particular one observes from
Fig.~\ref{Fig:Several-DRC} that in the case of $\ell=10^{-2}\lambda$ a
specular peak is hard to define at all in the mean DRC spectra.  This
is a clear indication of a highly rough surface and thus a very severe
test of our theory.

To further compare the analytic results derived earlier with those
obtained from the numerical simulation approach, we in
Fig.~\ref{Fig:Amplitude-Width} have plotted the amplitude of the
specular peaks~(circles) $\left.\left<\partial
  R_s/\partial\theta\right>\right|_{\theta=\theta_0}$, and their
width~(squares) $w(H,\ell/\lambda,\theta_0)$, as obtained from the
numerical simulation results shown in Figs.~\ref{Fig:Several-DRC}. 
The solid lines of this figure are the analytic
predictions for these quantities as given respectively by
Eqs.~(\ref{amplitude}) and (\ref{exactwidth}). As can be seen from this
figure, the analytic predictions are in excellent agreement with their
numerical simulation counterparts. In particular this confirms the
decaying and increasing power-laws in topothesy of exponent $1/H-1$ for
these two quantities respectively.

\begin{figure}[t!]
  \begin{center}
    \epsfig{file=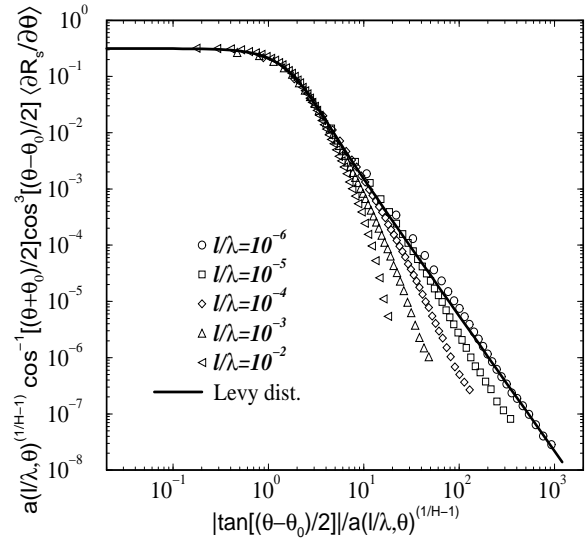,width=7.5cm,height=7.5cm}
  \end{center}
  \caption{Rescaled version of the rigorous numerical simulation
    results shown in Fig.~\protect\ref{Fig:Several-DRC}b. Only the
    data corresponding to $\theta<50^\circ$ are included.  In the
    rescaled coordinates all data~(symbols) should within the single
    scattering approximation collapse onto a L\'evy distribution of
    order $2H$ (solid line).}
  \label{Fig:Master}
\end{figure}

From Eqs.~(\ref{final-result}) we observe that if we replot the mean
DRC times the inverse of the prefactor of the L\'evy distribution, vs.
its argument, all mean DRC-curves corresponding to the same Hurst
exponent should (within single-scattering) collapse onto one and the
same master curve. This master curve should be the L\'evy
distribution, ${\cal L}_{2H}(x)$, of order $2H$.  Notice that this
data-collapse should hold true for arbitrary values for the angle of
incidence and topothesy.  The failure of such a data-collapse (onto
${\cal L}_{2H}$) indicates essential contributions from
multiple-scattering effects. The range of scattering angles where such
processes are important can therefore be read off from such a plot.
Furthermore, since the tails of the L\'evy distribution ${\cal
  L}_{2H}(x)$ drops off like $x^{-2H-1}$ (cf.  Eq.~(\ref{Levyinf}))
such rescaled mean DRC plots can be used to measure the Hurst exponent
of the underlying self-affine surfaces for which the light scattering
data have been obtained.  In order to check these predictions for our
numerical simulations results, we  present in
Fig.~\ref{Fig:Master} such a rescaling of the data originally
presented in Fig.~\ref{Fig:Several-DRC}b.  Only data lying to the left
of the specular peak have been included, {\it i.e.} only data for
scattering angles $\theta<\theta_0=50^\circ$.  As can be seen from
this figure the various scattering curves nicely fall onto the
master-curve (solid line) in regions where single-scattering is
dominating.  When

\noindent 
multiple scattering processes start giving a
considerable contribution the scattering curves start to deviate from
this master-curve. This observation could be used in practical
applications to determine for what regions the scattering is dominated
by single scattering processes. For the lowest topothesy considered
here, $\ell=10^{-6}\lambda$, a power-law extends nicely over large
regions of scattering angels --- a signature of the diffuse scattering
from self-affine surfaces.  According to Eq.~(\ref{diffuse-behavior})
the exponent of this power-law should be $-1-2H$.  A regression fit to
the scattering curve corresponding to the topothesy
$\ell=10^{-6}\lambda$ gives $H=0.73\pm0.02$, where the error indicated
is a pure regression error. The real error is of course larger.  With
the knowledge of the Hurst exponent obtained from the decay of the
diffuse tail of the mean DRC, we might now, based on the amplitude of
the specular peak, obtain an estimate for the topothesy of the
surface.  From the numerical simulation result we have that $\left.
  \left< \partial R_s/\partial \theta \right>
\right|_{\theta=\theta_0}\simeq 17.9$ which together with
Eq.~(\ref{amplitude}) gives $\ell=0.97 \cdot 10^{-6}\, \lambda$, where
we have used the value found above for the Hurst exponent.  These two
results fit quite nicely with the values $H=0.7$ and
$\ell=10^{-6}\lambda$ used in the numerical generation of the
underlying self-affine surfaces.

It should be noticed that for the numerical results presented in this
paper, we have not considered topothesies smaller then
$\ell=10^{-6}\lambda$.  However, since lowering the topothesy will, as
also indicated by our numerical results, favor single-scattering
processes over those obtained from multiple scattering, the analytic
results~(\ref{final-result}) will trivially be valid for low values of
the topothesy. This has also been checked explicitly by numerical
simulations (results not shown).

\begin{figure}[t!]
  \begin{center}
    \epsfig{file=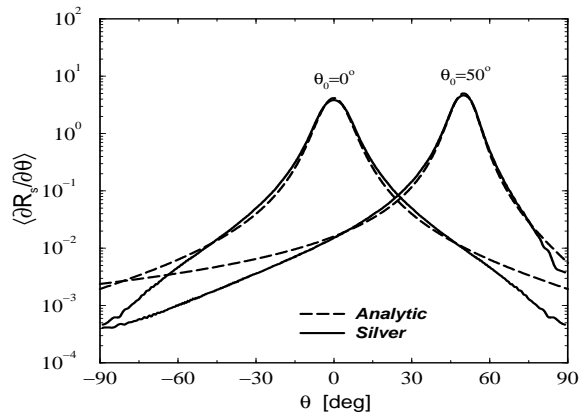,width=7.5cm,height=5.5cm}
  \end{center}
  \caption{The same as Fig.~\protect\ref{Fig:Compare}b, but now
    using a real metal~(silver) instead of a perfect conductor.  The
    value of the dielectric constant of silver at the wavelength of
    the incident light ($\lambda=612.7{\rm nm}$) was
    $\varepsilon(\omega) =-17.2+0.50i$.}
  \label{Fig:RealMetal} 
\end{figure}

So far in this paper we have assumed that the metal was a perfect
conductor. Obviously this is an idealization, and even the best
conductors known today are not perfect conductors at optical
wavelengths. By relaxing the assumption of the metal being perfectly
conducting to instead being a good conductor, {\it i.e.} a real metal,
we are no longer in position to obtain a closed form solution of the
scattering problem, the reason being that the boundary
conditions are no longer local quantities. In this latter case we
therefore have to resort to numerical calculations in any case.  In
order to see how well our analytic results~(\ref{final-result})
describe the scattering from real metals (in contrast to perfect
conductors) we in Fig.~\ref{Fig:RealMetal} give the mean DRC, as
obtained from numerical simulations~\cite{AnnPhys}, for a self-affine
silver surface of Hurst exponent $H=0.7$ and topothesy $\ell=10^{-4}
\lambda$. We recall that this choice for the topothesy corresponds to
a rather rough surface where the {\sc RMS}-height measured over the
whole length of the surface is $\sigma(L)\sim 1.45\lambda$.
Furthermore, the angles of incidence were $\theta=0^\circ$ and
$50^\circ$ and the wavelength of the incident light was
$\lambda=612.7~{\rm nm}$.  At this wavelength the dielectric constant
of silver is $\varepsilon(\omega) =-17.2+0.50i$~\cite{Palik}.  The
long dashed lines of Fig.~\ref{Fig:RealMetal} represent the
predictions from Eqs.~(\ref{final-result}), and as can be seen from
this figure, the correspondence is rather good. It is interesting to
see that the agreement between the analytical and numerical results is
of the same quality as the one found for the perfect conductor (see
Fig.~\ref{Fig:Compare}b). This indicates that the analytic results
given by Eqs.~(\ref{final-result}) are rather robust and tend to also
describe well the scattering from a good, but not necessary perfect,
reflector.  Simulations equivalent to those reported for silver have
also been performed for aluminum (results not shown) which has a
dielectric function that is more then three times higher at the
wavelength ($\lambda=612.7{\rm nm}$) used here. The conclusions found
above for silver also hold true for aluminum.  We find it interesting
to note that such self-affine aluminum surfaces were recently reported
to be seen for cold rolled aluminum~\cite{laminage}. The Hurst
exponents were measured to be $H=0.93\pm 0.03$ and $H=0.50\pm0.05$ for
the transverse

\noindent
and longitudinal direction, respectively.
Before closing this section it ought to be mentioned that for real
metals the numerical simulations approach based on Eq.~(\ref{int-eq}),
and used above, can no longer be used directly.  Instead a coupled set
of inhomogeneous Fredholm integral equations of the second type have
to be solved for the electric field, which is non-zero on the surface
of a real metal, and its normal derivative divided by the dielectric
constant of the metal.  Details about this approach can be found in
{\it e.g.}  Ref.~\cite{AnnPhys}.

\section{Conclusions}

We have considered the scattering of $s$-polarized plane incident
electromagnetic waves from randomly rough self-affine metal surfaces
characterized by the roughness exponent, $H$, and the topothesy,
$\ell$ (or slope $s(\lambda)$).  By considering perfect conductors, we
derive within the Kirchhoff approximation a closed form solution for
the mean differential reflection coefficient in terms of the
parameters characterizing the rough surface --- the Hurst exponent and
the topothesy (or slope) --- and the wavelength and the angle of
incidence of the incident light. These analytic predictions
(written from  a L\'evy distribution of index $2H$) were compared
against results obtained from extensive, rigorous numerical
simulations based on the extinction theorem. An excellent agreement
was found over large regions of parameter space. Finally the analytic
results, valid for perfect conductors, were compared to numerical
simulation results for a (non-perfectly conducting) aluminum
self-affine surface.  It was demonstrated that also in this case the
analytic predictions gave quite satisfactory results even though
strictly speaking they were outside their region of their validity.

\acknowledgements

The authors would like to thank Jean-Jacques Greffet,
Tamara Leskova and Alexei A. Maradudin for useful comments about this
work.  I.S. would like to thank the Research Council of Norway
(Contract No. 32690/213), Norsk Hydro ASA, Total Norge ASA and the
CNRS for financial support.


%
%
%
%
%
%
%
%
%
%
%
%
%

\end{document}